\documentclass[a4paper,twocolumn,10pt,accepted=2020-12-08]{quantumarticle}
\pdfoutput=1

\usepackage{graphicx}
\usepackage{hyperref}
\usepackage{float}
\usepackage{amsmath,amssymb,amsthm,amsfonts,amstext}
\usepackage{url}
\usepackage{color}
\usepackage{braket}
\usepackage{titlesec}
\usepackage{soul}
\usepackage[numbers, sort]{natbib}

\def\be{\begin{equation}}
\def\ee{\end{equation}}
\def\bea{\begin{eqnarray}}
\def\eea{\end{eqnarray}}
\def\bma{\begin{mathletters}}
\def\ema{\end{mathletters}}

\def\q0{\underline{0}}

\def\C{{\mathbb C}}
\def\id{{\mathbb I}}

\def\R{\mathbb{R}}

\def\one{\leavevmode\hbox{\small1\normalsize\kern-.33em1}}

\newcommand{\swap}{\mbox{SWAP} }
\newcommand{\swaps}{\mbox{SWAP}s }

\def\proj#1{\ket{#1}\!\bra{#1}}

\def\id{{\mathbb I}}

\renewcommand{\eqref}[1]{Eq.(\ref{#1})}
\newcommand{\figref}[1]{Fig.\ref{#1}}

\newcommand{\appref}[1]{App.\ref{#1}}

\newcommand{\miguel}[1]{{\color[rgb]{0.1,0.5,0.1}{#1}}}

\begin{document}

\title{Translating Uncontrolled Systems in Time}
\author{David Trillo, Benjamin Dive, and Miguel Navascu\'es}
\affiliation{Institute for Quantum Optics and Quantum Information (IQOQI), Austrian Academy of Sciences,\\
Boltzmanngasse 3, Vienna 1090, Austria}

\begin{abstract}
We show that there exist non-relativistic scattering experiments which, if successful, freeze out, speed up or even reverse the free dynamics of any ensemble of quantum systems present in the scattering region. This ``time translation'' effect is universal, i.e., it is independent of the particular interaction between the scattering particles and the target systems, or the (possibly non-Hermitian) Hamiltonian governing the evolution of the latter. The protocols require careful preparation of the probes which are scattered, and success is heralded by projective measurements of these probes at the conclusion of the experiment. We fully characterize the possible time translations which we can effect on multiple target systems through a scattering protocol of fixed duration. The core results are: a) when the target is a single system, we can translate it backwards in time for an amount proportional to the experimental runtime; b) when $n$ targets are present in the scattering region, we can make a single system evolve $n$ times faster (backwards or forwards), at the cost of keeping the remaining $n-1$ systems stationary in time. For high $n$ our protocols therefore allow one to map, in short experimental time, a system to the state it would have reached with a very long unperturbed evolution in either positive or negative time.
\end{abstract}

\maketitle

Since the industrial revolution, discerning which actions can speed up, decelerate or invert the natural evolution of physical processes, like chemical reactions, has been an overarching theme. Transformations mapping a physical system to some other point on its free evolution curve are known as \emph{time translations} \cite{sandu}. In quantum theory, the effect of a time translation over a system in state $\ket{\psi_0}$ with free Hamiltonian $H_0$ is to propagate the latter to $e^{-iH_0T}\ket{\psi_0}$, for some real $T$. For $T>0$, achieving such a transformation in time $T'=T$ just amounts to waiting for time $T$. The interesting time translations are those which can be completed in time $T'\not= T$.

There exist several mechanisms to conduct non-trivial time translations on a physical system. Some of them, based on quantum information processing, require the experimenter to possess a great deal of knowledge about the target system. Consider the simplest scheme, consisting in implementing the unitary $e^{-iH_0T}$ through a sequence of controlled operations. For a system of Hilbert space dimension $d$, this scheme requires knowing the $d^2$ real parameters determining the free Hamiltonian $H_0$ of the target. In addition, we need to know how the system responds to our attempts to control it, e.g., that if we generate a magnetic pulse along the $X$ axis, then the system will evolve according to the Pauli unitary $\sigma_x$. This amounts to specifying further $O(d^2)$ real parameters. Notably, there exist quantum processing protocols to reverse the time evolution of a system (i.e., to implement $e^{-iH_0T}$, with $T<0$) which do not demand the full knowledge of $H_0$ \cite{refocusing, spinEcho, dynamicalDecoupling,tokio}. All such protocols assume, however, knowledge of the remaining $O(d^2)$ continuous parameters describing our interaction with the target. That is, they demand full control of the system under consideration.

There exist other protocols which, for a given external observer, result in the same effects as a time translation. Perhaps the best known example of this is time dilation in special relativity: relative to a fixed observer, a system whose center of mass is made to move at relativistic speed for a time $T'$ will have its internal degrees of freedom time-translated by $T = T'/\gamma$ (where $\gamma$ is the Lorentz factor) rather than by $T'$, as it would have had the system been kept still. This can be used to realize translations of the form $0 < T \le T'$, but cannot be used to obtain a negative-time evolution. General relativistic effects could be used to achieve such effective negative-time translations of the internal degrees of freedom on a system, by moving it across a closed time-like curve \cite{goedel,thorne}. However, such schemes rely on some sort of natural time machines, whose existence is speculative and generally problematic \cite{Krasnikov2018}. Other methods, like the time translator of Aharonov et al. \cite{sandu}, while theoretically feasible, would operate under an astronomically small probability of success.

Here, we consider the problem of devising a realistic scheme to conduct general time translations over a target system with minimal assumptions on the latter. To this effect, we propose a class of non-relativistic scattering experiments. In these experiments, a number of particles are produced, allowed to propagate without intervention on a path which may or may not interact with the target system, and subsequently measured after some time $T'$. Depending on the outcome of this measurement, the experiment is regarded as either a ``success'' or a ``failure''. When the scattering region holds $n$ identical quantum systems and the experiment succeeds, then each system $i$ is guaranteed to leap to the quantum state it would have had if it had been evolving unperturbed for time $T_i$, where $T_i$ can be different from $T'$, and even negative. The experiments do not rely on any knowledge of the Hamiltonian of the target systems or their interaction with the scattered particles, depending only on $d$ and the desired $T_i$. For high $n$, some of these experiments can quickly propagate one of the systems to its far past or future, at the cost of freezing the rest in time. For $n=1$ our results resonate with those of \cite{resetting}, where a similar setup was used to send a system back to the state it was in before the experiment started. In this regard, our present work shows that a single system can be rewinded much faster than the protocols introduced in \cite{resetting} allowed. 

Compared to the aforementioned protocols \cite{goedel, thorne, sandu}, we do not require any relativistic or gravitational effects. Furthermore, in contrast to the quantum information processing protocols \cite{refocusing, spinEcho, dynamicalDecoupling,tokio}, where each step depends on the $O(d^2)$ continuous parameters specifying our interaction with the system, our scheme just depends on one natural number: the Hilbert space dimension of the target. Since non-relativistic time translation is impossible without this knowledge\footnote{To see this, take {\cal P} to be a heralded protocol based on quantum probes that translates system $S$, with non-trivial free Hamiltonian $H_S$, by a non-trivial amount. No matter what {\cal P} is, a larger system can be constructed which it will be unable to translate in time. Consider a system $R$, with non-trivial free Hamiltonian $H_{R}$, that does not interact at all with the probes used by {\cal P}. If {\cal P} acts on the higher-dimensional joint system $SR$, with free Hamiltonian $H_S\otimes \id_R+\id_S\otimes H_R$, it will still flag “success” with the same probability. However the joint system has not undergone a consistent time translation, as $S$ will be translated by the original amount while $R$ evolves freely.}, our methods are as universal as the laws of quantum mechanics permit.

\section*{The model}

\begin{figure}
  \centering
  \includegraphics[trim=3cm 2cm 3cm 2cm, clip, width=0.95\columnwidth]{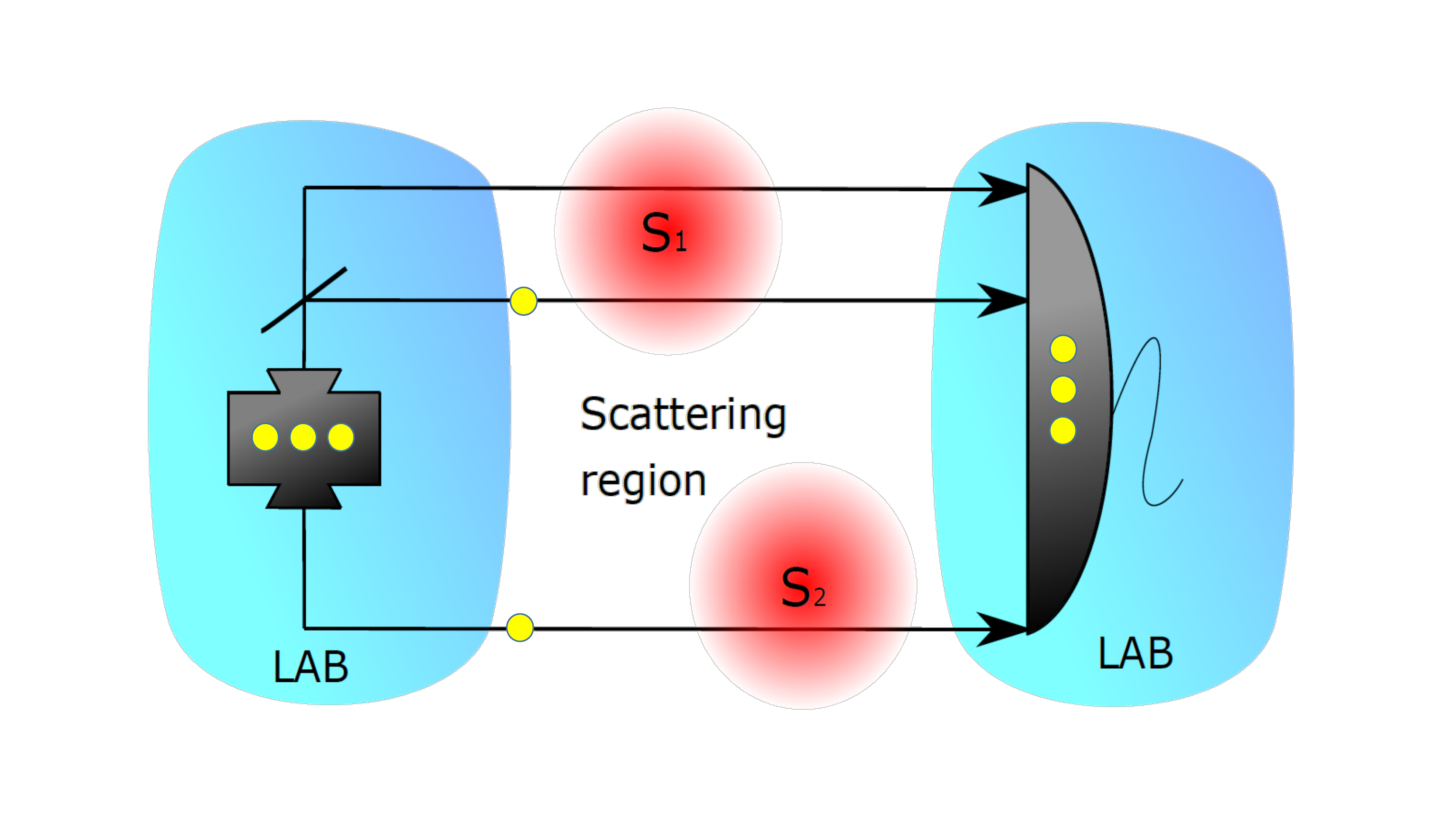}
  \caption{\textbf{The setup.} The large red discs labelled $S_1$ and $S_2$ are the systems and their interaction radius, which are well separated to ensure that the probes (small yellow discs) only interact with one at a time. The left section is the `preparation' part of the the lab, where the probes are prepared in a desired quantum state. After scattering with one of the systems, these probes are then measured (in the right section) in order to herald a successful run of the experiment.}
  \label{modelPic}
\end{figure}

We consider a scenario where the experimental setup consists of two parts: a \emph{controlled lab}, where we can prepare any quantum state and conduct any quantum operation; and a \emph{scattering region}. The latter contains $n$ identical physical \emph{target systems} of Hilbert space dimension $d$ at separate locations, see \figref{modelPic}. We assume that they remain in the same place during the course of the experiment. The initial (internal) quantum state of the $n$ systems is unknown; for simplicity, we will take it to be pure and denote it by $\ket{\psi_{1,...,n}}$. 

If left unperturbed, each of these systems will independently evolve according to an (unknown) time-independent Hamiltonian $H_0$. That is, after time $T$ the state of the $n$ systems will evolve to $e^{-i\sum_{k=1}^nH^{(k)}_0T}\ket{\psi_{1,...,n}}$. To incorporate decay processes, we allow $H_0$ to be non-Hermitian. 

To influence systems $k=1,...,n$, we can prepare a particle in the controlled lab and let it propagate within the scattering region. While in the scattering region, these particles or \emph{probes} interact with each system $k$ via the Hermitian or non-Hermitian Hamiltonian $H_I(\vec{r}-\vec{q}_k)$, where $\vec{r}$ ($\vec{q}_k$) denotes the probe's (system $k$'s) position. The joint state of systems $1,...,n$ and $P$ thus evolves as

\begin{align}
&i\frac{\partial}{\partial t} \ket{\psi_{1,..,n, P}} = \big(H_P+\sum_{k=1}^n H^{(k)}_0+H^{(k,P)}_I\big)\ket{\psi_{1,..,n,P}},
\end{align}
\noindent where $H_P$ denotes the free Hamiltonian of the probe. For technical reasons, $H_I$ is assumed to be a bounded operator; otherwise $H_I$, $H_0$ and $H_P$ are arbitrary and unknown. The fact that $H_I$ is unknown is why we call the system \emph{uncontrolled}, as it means that we are unable to tune any experimental degree of freedom such that it directly impacts its dynamics in a known way.

We will allow multiple probes at a time within the scattering region. This could give rise, in principle, to complex many-body interactions. Notwithstanding, we will impose the additional condition, which we will call the \emph{targeting assumption}, that it is possible to prepare a probe in such a way that it interacts with a single uncontrolled system (its target) and nothing else. Such a probe will either return to the lab within a given time $\Delta t$ and through a given channel, or else be absorbed by the environment or lost in free space. Moreover, if several probes with different targets are prepared simultaneously, then the evolution of each probe and its target will be independent and identical among the different pairs of probe-uncontrolled system. Meanwhile, those uncontrolled systems without a targeting probe will keep evolving through $H_0$. This targeting assumption can be justified in many experimental setups where the uncontrolled systems are sufficiently separated in space.

In our protocols, we allow each probe to enter the scattering region and return to the controlled lab repeatedly. While in the lab, we are allowed to interact with the probe in any controlled way. We can, e.g. entangle it with a quantum memory inside the lab, or with any other probe that happens to be within the lab at the same time. Note that, except for the presence of several uncontrolled systems/probes in the scattering region, this is similar to the scenario considered in \cite{resetting}. 

The scattering protocol ends at time $T'$, when we conduct a dichotomic heralding measurement over the quantum registers present in the lab. If the outcome is ``success'', we desire the state of systems $1,...,n$ to be
\be
\ket{\psi'_{1,...,n}}=U(T_1,...,T_n)\ket{\psi_{1...,n}},
\label{resultado}
\ee
with $U(T_1,...,T_n)=\bigotimes_{k=1}^n e^{-iH_0T_k}$. 

Throughout the text, we will further assume that the time $\Delta t$ that each probe spends interacting with a target can be taken arbitrarily small. This condition is not necessary to devise sound time translation protocols, and, in fact, all our protocols work for finite $\Delta t$. However, to obtain an overall bound for the minimum runtime $T'$ needed to effect a desired translation $T_1,...,T_n$, will require taking the limit $\Delta t\to 0$.

The main focus of this paper is on characterizing what time translation experiments are possible with a non-zero probability of success; calculating the general probability for an arbitrary $n$, $d$, and target time translation, is beyond the present scope. This is because the general protocols we give in our constructive proof are highly inefficient and the success rate itself depends on the free and interaction Hamiltonians. Therefore, we do not know of any general scaling law for success in terms of $n$ or $d$. For particular small protocols we are able to optimize the method for Haar random Hamiltonians through a mix of analytic and numerical methods discussed later. Thus, we have that, on average, rewinding a single qubit can be done with a probability of $5\%$, while fast forwarding one qubit in an ensemble of two has a $0.02\%$ chance of success. While these numbers are not large, there is no reason to believe that protocols with a higher success rate do not exist, perhaps at the cost of additional complexity.

\section*{Results}

The main result of our work is that scattering experiments of the type described above, of duration $T'$, leading to time translations of the form of \eqref{resultado}, are possible if and only if 
\be
\sum_{i:T_i>0} T_i+\sum_{i:T_i<0} |T_i|(d-1)\leq nT'.
\label{general}
\ee
We stress that, as required by the scenario we are considering, this is independent of $H_0$, $H_I$, and the initial state of the target systems. In the next section, we prove that it is impossible to implement any other set $\{T_i\}_i$, and then describe in the following one how to construct a scattering experiment that realizes a valid set of time translations. Before doing this, we now discuss the interpretation of \eqref{general} and its implications.

For $n=1$, it simplifies to $T_1\in [-\frac{T'}{d-1},T']$. This means that, in principle, we could invert the evolution of the uncontrolled system in the scattering region.  This is similar to what was done in \cite{resetting}; however, the protocol described there required an experiment of duration $T'=O(d^2)|T_1|$, as opposed to $(d-1)|T_1|$. The latter bound is consistent with the work of \cite{tokio}, where the authors prove that, in order to invert a unitary probabilistically (in a controlled system), at least $d-1$ uses thereof are needed.

For $n=1$, \eqref{general} also implies that one cannot \emph{fast-forward} the uncontrolled system. That is, in order to effect the transformation $\ket{\psi_1}\to U(T_1)\ket{\psi_1}$, with $T_1>0$, then one needs to invest, at least, time $T_1$. That fast-forward is impossible for unknown Hamiltonians was already established in \cite{FF,FF2,noClonU} (note that fast-forward is possible, though, for controlled systems under the promise that the Hamiltonian belongs to a family of quadratic polynomials of the canonical operators \cite{Burd2019}). In this regard, we show that such no-go results also extend to scenarios where we allow for arbitrarily small probabilities of success.

Fast-forwarding is compatible with \eqref{general} only when there is more than one uncontrolled system within the scattering region, such as with configuration $T_1=nT',T_2=...=T_n=0$. This opens the door to projecting a single uncontrolled system to its far future at the cost of freezing the evolution of the rest during the scattering process. A second interesting configuration, that we call fast rewinding, is $T_1=-\frac{nT'}{d-1}, T_2=...=T_n=0$. 

More generally, \eqref{general} is equivalent to the following postulates: a) evolution time cannot be created; b) evolution time can be transferred between two identical systems at no cost; c) evolution time within a system can be inverted at a cost $(d-1)$, where $d$ is its Hilbert space dimension; d) evolution time can be destroyed. 

\section*{Maximal rate of time translations}
We now prove that it is impossible to realize time-translations that do not obey \eqref{general} using the general scattering experiments introduced previously. To do this, we first identify the form of the most general transformation that we can effect within this model.

When a probe or collection thereof interacts with an uncontrolled target system from time $t=T_0$ to $t=T_1$, the Hamiltonian guiding the evolution of the joint system is:
$$
H_0+\tilde{H}_P+\tilde{H}_I,
$$
where $\tilde{H}_P,\tilde{H}_I$ are, respectively, $\sum_{\kappa=1}^N H_P^{(\kappa)}$ and $\sum_{\kappa=1}^N H_I^{(\kappa)}$. If $H_I$ is bounded and the number of probes $N$ is finite, then we can use the Dyson series in the Schrödinger picture to model the evolution of the joint system as $\ket{\psi}\to\tilde{U}(T_1,T_0)\ket{\psi}$, with $\tilde{U}(T_1,T_0)$ given by
\begin{align}
&\sum_{j=0}^\infty (-i)^j\int_{T_1 \geq t_1\geq ...\geq t_j\geq T_0}dt_1\cdots dt_j \nonumber\\
&e^{-i(\tilde{H}_P+H_0)(T_1-t_1)}\tilde{H}_Ie^{-i(\tilde{H}_P+H_0)(t_{1}-t_2)}\tilde{H}_I...\nonumber\\
&e^{-i(\tilde{H}_P+H_0)(t_{j-1}-t_j)}\tilde{H}_Ie^{-i(\tilde{H}_P+H_0)(t_j-T_0)}.\nonumber
\end{align}
Whatever post-selection we might be applying to the probes will just affect system $1$ through the bipartite terms $\tilde{H}_I$ in the expression above. Taking into account that $e^{-i(\tilde{H}_P+H_0)t}=e^{-i\tilde{H}_Pt}e^{-iH_0t} = e^{-iH_0t}e^{-i\tilde{H}_Pt}$, we may group the terms of the form $e^{i\tilde{H}_Pt_k}\tilde{H}_I e^{-i\tilde{H}_Pt_{k+1}}$ and so after measuring we are left with a continuous linear combination of terms of the form
\be
\Pi_{1}e^{-iH_0t_1}\Pi_{2}e^{-iH_0t_2}...,
\label{decomp}
\ee
where $\sum_i t_i=T_1-T_0$, and the $\Pi_{j}$ depend on the Hamiltonians $\tilde{H}_I, \tilde{H}_P$ and the measurement outcome. 

In general in a protocol we take the segment $[0, T']$ and choose some sequence of times $0 = T_0 < T_1 < ... < T_M = T'$. Between times $T_i$ and $T_{i+1}$ we can decide whether we send some probes to interact with the target system of not. If we do not, the target system evolves via the operator $e^{-iH_0 (T_{i+1}-T_i)}$. Otherwise, it evolves via an operator that we can decompose as before in terms of the form of \eqref{decomp} . At the end of the protocol, after the post-selection, we thus have a linear combination of terms of the form of \eqref{decomp}, but with $\sum_i t_i = T'$.

Therefore, the target system will have a state proportional to
$$
 \sum_j A_j\proj{\psi_{1}} A_j^\dagger,
$$
where $\ket{\psi_{1}}$ is the initial state of the target, the index $j$ labels the measurement outcomes that we post-select to and $A_j$ is an operator which can be expressed as a continuous linear combination of terms of the form of \eqref{decomp} with  $\sum_i t_i = T'$

Suppose now that we have one target, and we are interested in effecting the transformation $\ket{\psi_{1}} \mapsto e^{iH_0 T} \ket{\psi_1}$. We must therefore have
$$
\sum_j A_j\proj{\psi_{1}} A_j^\dagger\propto e^{-iH_0T_1}\proj{\psi_{1}} e^{iH_0T_1},
$$
with $T_1\equiv -T<0$. It follows by convexity that the above equation can hold for some non-zero proportionality scalar only when all the non-zero $A_j$'s are proportional to $e^{iH_0T}$. Let then $A$ be one of such non-zero terms $\{A_j\}_j$ corresponding to one possible measurement result of the probes. From the above, we have that
\be
A\propto e^{iH_0T}.
\label{time_warp}
\ee
\noindent for all Hamiltonians $H_0, H_I, H_C$.

Suppose then that there exists such an operator $A$, and let $H_0$ be any generic Hamiltonian such that the right hand side of \eqref{time_warp} does not vanish, for some fixed $H_I$, $H_P$. Since $H_0$ is a generic operator, it must admit a Jordan decomposition of the form $B^{-1}H_0B=\sum_{i=1}^d\alpha_i\proj{i}$, with $\{\alpha_k\}_{k=1}^d\subset \C$, for some invertible $d\times d$ matrix $B$. Since $A=f(\vec{\alpha})e^{iH_0T}$ for some scalar $f(\vec{\alpha})$, we have that

\be
\bra{j}\tilde{A} \ket{j} =f(\vec{\alpha})e^{i\alpha_jT},
\label{alfalfa}
\ee
for $j=1,...,d$, where here $\tilde{A}$ denotes $B^{-1}AB$. On the other hand, braketing \eqref{decomp} between $\bra{j}B^{-1} \bullet B \ket{j}$  we obtain
\[
\sum_{\ell_1, \ell_2, ...} \bra{j}\Tilde{\Pi}_1 \ket{\ell_1} \bra{\ell_1} \Tilde{\Pi}_2 \ket{\ell_2} \cdots \ e^{-i(\alpha_{\ell_1}t_1+\alpha_{\ell_2}t_2 + \cdots)},
\]
where we are again denoting $\tilde{\Pi}_i := B^{-1}\Pi_i B$. By grouping the terms in the exponent with the same $\alpha_i$ we may write each term as some coefficient times $e^{-i\vec{\alpha}\cdot \vec{t}}$, where now $\vec{t}$ lives in $\R^d$, but still satisfies $\sum_i t_i= T'$ and positivity. Therefore, the full linear combination can be written as
\be
\bra{j} \tilde{A} \ket{j}=\int_{{\cal T}}c^j(\vec{t})e^{-i\vec{\alpha}\cdot\vec{t}} d\vec{t},
\label{heno}
\ee
by choosing appropiate coefficients $c^j(\vec{t})$, and where ${\cal T}=\{\vec{t}\in \R^d, t_i\geq 0,\sum_i t_i= T'\}$.
From (\ref{alfalfa}) and (\ref{heno}), it follows that $f(\vec{\alpha})$ admits the decompositions
$$
f(\vec{\alpha})=\int_{ {\cal T}_j}\tilde{c}^j(\vec{t})e^{-i\vec{\alpha}\cdot\vec{t}}d\vec{t},
$$
for $j=1,...,d$. Here ${\cal T}_j=\{\vec{t}\in \R^{d}, t_i\geq T\delta_{ij},\sum_i t_i= T'+T\}$.

Now, express the vector $\vec{\alpha}\equiv \vec{\beta}+i\vec{\gamma}$ in terms of its real and imaginary parts $\vec{\beta},\vec{\gamma}$. Fixing $\vec{\gamma}$, we have that the above expressions depend on $\vec{\beta}$ as
$$
f(\vec{\alpha})=\int_{{\cal T}_j}\tilde{c}^j(\vec{t},\vec{\gamma})e^{-i\vec{\beta}\cdot\vec{t}}d\vec{t}.
$$
for some new coefficients $\tilde{c}$. In particular, for all $k, j \in \{1,...,d\}$,
$$
\int_{{\cal T}_j}\tilde{c}^j(\vec{t},\vec{\gamma})e^{-i\vec{\beta}\cdot\vec{t}} d\vec{t}=\int_{ {\cal T}_k}\tilde{c}^k(\vec{t},\vec{\gamma})e^{-i\vec{\beta}\cdot\vec{t}} d\vec{t}.
$$
\noindent This holds for all $\vec{\beta} \in \R^d$. Multiplying the above expression by $e^{i \vec{\beta}\cdot \vec{t}_0}$ for some $\vec{t}_0 $ and integrating with respect to $\vec{\beta}$ one can see that $\tilde{c}^j(\vec{t}_0,\vec{\gamma})$ must vanish if $\vec{t}_0\in {\cal T}_j \setminus {\cal T}_k$. Therefore, we may write
$$
f(\vec{\alpha})=\int_{\cap_j{\cal T}_j}\tilde{c}(\vec{t},\vec{\gamma})e^{-i\vec{\beta}\cdot\vec{t}}d\vec{t}.
$$
Any $\vec{t}\in \cap_j{\cal T}_j$ satisfies $t_i\geq T$ and $\sum_{i=1}^d t_i= T+T'$. Combining these two expressions, we conclude that, for all $\vec{\gamma}$ for which $f(\vec{\alpha})$ does not vanish, $dT\leq \sum_{i=1}^d t_i= T+T'$. That is,
\be
(d-1)|T_1|\leq T'.
\label{limitReset}
\ee
Note that the argument above does not invoke at any point the uncontrollability of system $1$: it holds even if we know the form of the operators $\{\Pi_i\}_i$ in \eqref{decomp}. In fact, it holds if we further know the similarity transformation that diagonalizes $H_0$. 

We can now build from this result to the general scenario. Suppose that, through a scattering experiment of duration $T'$, we were able to induce a transformation of the type $A\propto U(T_1,...,T_n) := \bigotimes_{j=1}^n e^{-iH_0T_j}$, for some times $T_1,...,T_n$. Let us assume, w.l.o.g., that $T_1,...,T_k<0$, and $T_{k+1},...,T_n\geq 0$. 

Now, let $B$ be the similarity transformation that diagonalizes $H_0$, i.e., $B^{-1}H_0 B=\sum_{i}\alpha_i\proj{i}$, and consider the operator 

$$
\tilde{A}:=\prod_{j=1}^{d-1}\left(B^{\otimes n}\Gamma^j(B^{-1})^{\otimes n}\right)A\left(B^{\otimes n}\Gamma^{-j} (B^{-1})^{\otimes n}\right), 
$$
with $\Gamma=\id^{\otimes k}\otimes \tilde{\Gamma}^{\otimes n-k}$, where $\tilde{\Gamma}:=\sum_{i=1}^{d}\ket{i}\bra{i\oplus 1}$ and $\oplus$ is addition modulo $d$. Noting that, for any $d\times d$ diagonal matrix $Y$, $\prod_{j=1}^{d-1} \tilde{\Gamma}^j Y\tilde{\Gamma}^{-j}=\mbox{det}(Y)Y^{-1}$, we have that $\tilde{A}\propto U((d-1)T_1,...,(d-1)T_k,-T_{k+1},...,-T_n)$.

Finally, define the linear map $\Lambda:M_{d}^{\otimes n}\to M_{d}$ by 
\begin{align}
\Lambda(X):= \sum_{i_1,...,i_{n-1}}&\left(\id\otimes \bra{i_1}\otimes \cdots \otimes \bra{i_{n-1}}\right) X \nonumber \\
&\qquad\qquad \left(\ket{i_1}\otimes \cdots \otimes \ket{i_{n-1}}\otimes\id\right). \nonumber
\end{align}

This map implements the linear extension of the operation 
$$
\Lambda(X_1\otimes \cdots \otimes X_n)=X_1 \cdots X_n,
$$
so it follows that 
$$
\Lambda(\tilde{A})\propto U\left((d-1)\sum_{i=1}^k T_i-\sum_{i=k+1}^n T_i\right),
$$
i.e., $\Lambda(\tilde{A})$ is a rewinding transformation for $1$ target system. Clearly, $\Lambda(\tilde{A})$ can be expressed as linear combinations of product operators of the form \eqref{decomp} for $n=1$, with the particularity that $\sum_i t_i= (d-1)nT'$. By \eqref{limitReset} we have, then, that the total rewinding time $(d-1)\sum_{i=1}^k |T_i|+\sum_{i=k+1}^n |T_i|$ is upper bounded by $nT'$. That is
$$
\sum_{i:T_i>0} T_i+\sum_{i:T_i<0} |T_i|(d-1)\leq nT'
$$ as stated in \eqref{general}.

\section*{Realizing arbitrary time translations}
\noindent\emph{Scattering experiments as matrix polynomials}

We now show that every time translation allowed by \eqref{general} can be asymptotically realized by a scattering experiment. The most natural and convenient way to do this is in terms of polynomials of matrix variables; we demonstrate how these arise from the setup considered above and show their equivalence here. We begin by restricting ourselves to a subset of possible scattering experiments which turn out to be sufficient to implement all time translations. 

Take first the case of a single target system. We prepare a probe in a superposition of states, one inside and another one outside the lab, controlled by an internal qubit register $R$. That is, the system register-probe is prepared in the state $\frac{1}{\sqrt{2}}(\ket{0}_{R}\ket{\Phi}_P+\ket{1}_{R}\ket{\varphi}_P)$, where $\ket{\varphi}$ is the state that allows the probe to interact with system $1$ or else be absorbed (which exists by virtue of the targeting assumption) and $\ket{\Phi}$ is some bound state within the lab. 

The world line marked by state $\ket{1}_R$ hence propagates through the scattering region, interacting with system $1$, until it re-enters the lab after time $\Delta t$. In that moment, we perform the operation
\be
\proj{0}_R\otimes \id_P + \proj{1}_R\otimes \ket{\Phi}\bra{\tilde{\varphi}},
\label{conditional}
\ee
\noindent where $\ket{\tilde{\varphi}}$ is any state with support within the lab. If $\ket{\psi_1}$ is the initial state of system $1$, then the final (unnormalized) joint state of register, target and probe is
$$
\frac{1}{\sqrt{2}}\left(\ket{0}_{R}\otimes V\ket{\psi_1}+ \ket{1}_{R}\otimes W\ket{\psi_1}\right)\otimes \ket{\Phi}_P,
$$
\noindent where $V=e^{-iH_0\Delta t}$ and 
\be
W = \left(\id_1\otimes\bra{\tilde{\varphi}}_P\right)  e^{-i(H_0+H_P+H_I)\Delta t}\left(\id_1\otimes\ket{\varphi}_P\right).
\ee
\noindent Since the final state $\ket{\Phi}$ of the probe plays no further role, in the following we omit it.

If we repeat the above steps sequentially at intervals of $\Delta t$ on the same system, then the final joint state of the lab's quantum memory and the target system is
$$
\frac{1}{\sqrt{2^m}}\sum_{j_1,...,j_m=0,1}\ket{j_1,...,j_m}_R\otimes X_{j_m}...X_{j_1}\ket{\psi_1},
$$
where we have used the shorthand $X_0 = V$ and $X_1 = W$.  If we now post-select the quantum memory to the pure state $\sum_{\vec{\jmath}}g^*_{\vec{\jmath}}\ket{j_1,...,j_m}$, the final state of system $1$ is
\be
\label{eq:matpol}
\sum_{j_1,...,j_m=0,1}\frac{g_{j_1,...,j_m}}{\sqrt{2^m}} X_{j_m}...X_{j_1} \ket{\psi_1}= \frac{G(V,W)}{\sqrt{2^m}} \ket{\psi_1},
\ee
where $G(V,W):=\sum_{\vec{\jmath}}g_{\vec{\jmath}}X_{\vec{\jmath}}$ is a \emph{homogeneous matrix polynomial of degree $m$} \cite{formanek}.

In other words, the result of the scattering experiment is fully described by the matrix polynomial $G(V, W)$, which is determined by the state  $\sum_{\vec{\jmath}}g^*_{\vec{\jmath}}\ket{j_1,...,j_m}$ we post select the quantum memory to. This polynomial expresses how the system state will evolve between the start and end of the experiment as a function of the unknown propagators $V$ and $W$ (themselves functions of the unknown Hamiltonians). Conversely, given any homogeneous matrix polynomial $G(V, W)$, there exists a scattering experiment that propagates the target system by $G(V,W)$ (up to normalization). In such a scattering experiment, the number of probes corresponds to the degree of the polynomial; and the coefficients of the vector that post-selects the quantum memory ($g^*_{\vec{j}}$) to the complex conjugate of the coefficients of the matrix polynomial itself. Note also that the probability of success of the experiment is given by the norm squared of the final state of \eqref{eq:matpol}. In the case that $G(V, W)$ is a time translation of the form of \eqref{resultado}, or more generally is proportional to a unitary, this depends on the unknown variables $V, W$ but not on the initial system state $\ket{\psi_1}$.

In the case of $n$ identical target systems, everything follows in much the same way. For the $k^\text{th}$ system, we use the targeting assumption to sequentially prepare $m$ probes, each in a superposition of states, conditioned on the quantum memory in the lab, which either stay completely within the lab or interact only with system $k$. On re-entering the lab, each of these probes and their corresponding registers are subject to the gate (\ref{conditional}). The process just described is conducted simultaneously on all target systems $1,...,n$. After time $T'=m\Delta$, when all $nm$ probes have returned to the lab, the joint state of the quantum memory and the $n$ target systems is thus
$$
\frac{1}{\sqrt{2^{m n}}}
\sum_{\small \substack{\vec{\jmath}^{\;1},...,\vec{\jmath}^{\;n}\\\in(0, 1)^{m}}}
\ket{\vec{\jmath}^{\;1}}...\ket{\vec{\jmath}^{\;n}} \otimes \big(X_{\vec{\jmath}^{\;1}}  \otimes...\otimes X_{\vec{\jmath}^{\;n}}\big) \ket{\psi_{1...n}}.
$$
Now post-selecting the $n m$ qubit quantum memory to the pure state
$$
\sum_{\vec{\jmath}^{\;1}, ..., \vec{\jmath}^{\;n}} g^*_{\vec{\jmath}^{\;1}, ..., \vec{\jmath}^{\;n}} \ket{\vec{\jmath}^{\;1}}...\ket{\vec{\jmath}^{\;n}}
$$
leaves the $n$ target systems in the final state $\tfrac{1}{\sqrt{2^{m n}}} G(V, W) \ket{\psi_{1...n}}$ where
\begin{equation}
G(V, W) :=  \sum_{\vec{\jmath}^{\;1}, ..., \vec{\jmath}^{\;n}} g_{\vec{\jmath}^{\;1}, ..., \vec{\jmath}^{\;n}}
 X_{\vec{\jmath}^{\;1}} \otimes ... \otimes X_{\vec{\jmath}^{\;n}}
\end{equation}
We stress that, while the notation is unfortunately messier, there is no major conceptual difference between the single particle case and the multipartite one. In the latter, we call $G(V, W)$ a \emph{homogeneous tensor polynomial of degree $m$ in $n$ parties}. This describes how the system states evolve between the start and end of experiment as a sum over terms, each of which has $n$ tensor products which themselves each contain $m$ dot products of $V$ or $W$. As before, the equivalence between scattering experiments and $G(V, W)$ is the same, and the probability of success can be extracted in the same way.

In everything that we have considered here our matrix polynomials have only two unknowns, namely $V$ and $W$. For our more general constructs it will be necessary to have a larger number; this can be done in several ways. One is to consider a coarse-graining of time scales where several probes are treated as one, and the corresponding elements of the quantum memory are treated as a single higher dimension qudit. In this way we would map $V ... V V \to V$, $V ... V W \to W_1$, $V ... W V \to W_2$ and so on, which allows us to have access to arbitrarily many different matrix variables. An alternative is to consider a larger number of possible paths per probe. Rather than just having one that corresponds to no scattering and another giving rise to the evolution operator $W$, we can instead consider $p$ different interacting paths leading to some different scattering propagators $W_p$. In this case, each probe would be initialized in an equal superposition of all possible paths under consideration, conditioned on the state of a $p+1$ dimensional qudit in the lab's quantum memory. Either methods allow us to define a scattering experiment for the homogeneous polynomials $G(V, W_1, W_2, ...) = G(V, \vec{W})$ in an identical way to before.

What we have provided here is therefore a 1:1 equivalence between a certain class of scattering experiments and homogeneous (tensor) matrix polynomials. Once we have found a polynomial that realizes a desired time-translation operation, we can directly read off a scattering experiment that would implement it. There are, however, more scattering experiments possible than the canonical class discussed here. For example, a different initial state could be used for the probes, or a higher rank post selection used on the quantum memory. In the following, we show that any feasible time translation can be implemented by means of a canonical scattering protocol. However, as we will see, in some cases there exist non-canonical scattering protocols which enable a higher probability of success.\\

\noindent\emph{Protocols in $n=1$}

Since fast-forwarding is impossible in the $n=1$ case, the only interesting time translation in this scenario is time reversal. In the language of matrix polynomials, that requires one with the property that $G(V,W)\propto V^{-s}$ for some fixed $s$, when evaluated on $d\times d$ matrices $V$ and $W$. In \appref{app:RewindingPoly}, we show how to construct such a homogeneous matrix polynomial, denoted by $R(V, \vec{W})$, with exactly this property. This polynomial has degree $m = s (d-1) + d^2$, where $d$ is the dimension of the target system; it can hence be implemented by means of a canonical scattering protocol in time
\begin{equation}
T' = m \Delta t = \Delta t s (d-1) + \Delta t d^2,
\end{equation}
and it translates the system in time by $T_1 = - s \Delta t$. Taking the limit of small time steps $\Delta t \to 0$ while proportionately increasing the number of steps $s$ leads to the quotient $\frac{T'}{|T_1|}$ approaching $d-1$ from below. This is precisely the limit of how well single-system time rewinding can be done according to \eqref{general}. From this, achieving any value of $T_1$ within the interval $\left(-\frac{T'}{d-1},T'\right)$ is simply a matter of applying the previous rewinding protocol for time $\alpha T'$, with $0\leq \alpha\leq 1$, and then letting the system evolve naturally for time $(1-\alpha)T'$.\\

It is illuminating to show exactly how this works for the simple case of a qubit. Consider the matrix polynomial $F(A, B)=[A, B]^2$, where $A$ and $B$ are $2\times 2$ matrices. As the commutator is traceless and the square of a traceless $2\times 2$ matrix is proportional to the identity, we have that $F(A, B)\propto\id_2$. Performing the substitution $A\to WV^s, B\to V$, we have that 
$$
F(WV^s, V) = [W,V]V^s[W,V]V^s \propto \id_2.
$$
Rearranging this easily yields
\be
R(V, W) := [W,V]V^s[W,V] \propto V^{-s}
\label{qubitRewinding}
\ee
which is precisely the matrix polynomial that we desire, with the optimal degree of $4+s$.

The probability of success of the canonical scattering experiment arising from this polynomial naturally depends on $V$ and $W$. If the two commute, for example, then it will clearly always fail as even though the results are technically correct, the constant of proportionality is $0$, but this only happens for a measure $0$ set of matrices $V$ and $W$. If, however, they are independently sampled from the Haar measure then the average probability of successful post-selection is around $0.05\times 2^{-s}$.

However this probability can be increased because, in this case, it is possible to construct far more efficient scattering protocols than the canonical one. In order to propagate system $1$ through $R(V,W)$, one can, e.g., propagate system $1$ by the polynomial $[W,V]$, then wait for time $s\Delta t$ and again propagate system $1$ by $[W,V]$. This would require using $4$ rather than $4+s$ probes; hence raising the probability of success to $5\%$, independently of $s$. Furthermore, ongoing work \cite{forthcoming} shows that there exist rewinding protocols for qubits with arbitrarily high average probability of success of running time $T'=T_1 + \epsilon$, with $\epsilon$ independent of $T_1$.\\

\noindent\emph{Protocols in $n=2$}

The two system case has a much richer structure of possible time translations. In order to show that every pair $\{T_1, T_2\}$ that satisfies \eqref{general} for $n=2$ can be reached asymptotically, it is sufficient to consider the extremal protocols $\{T_i = 2 T', T_j=0\}$ (optimal time transfer) and $\{T_i = -2 T'/(d-1), T_j = 0\}$ (fast rewinding) for $i\neq j \in \{1, 2\}$. Any other solution of \eqref{general} can be reached by sequentially applying optimal time transfer, fast rewinding and free evolution, in an analogous way as for $n=1$.

To build our intuition for how to do this we begin by noting that, if the two targets were fully controllable, then time transfer could be achieved via probabilistic teleportation. Indeed, let $\ket{\psi_{1,2}}$ be the initial state of systems $1, 2$ and suppose that we wished to make system $1$ evolve $2T'$ in time $T'$. We need to prepare two probes $A, A'$ in a maximally entangled state of the form $\ket{\Psi^+}_{AA'}$, perform a \swap operation between systems $A'$ and $2$, wait for time $T'$ and then swap the systems again. This will have the effect of propagating systems $1, A'$ by $e^{-iH_0T'}$ each. If we now project the bipartite systems $1, A$ to the maximally entangled state and then swap systems $A', 1$, then the state of systems $1,2$ will be $(e^{-2iH_0T'}\otimes \id)\ket{\psi_{1,2}}$.

Unfortunately, in our considered scenario, the systems are uncontrolled so we do not have the ability to directly implement a \swap between systems and probes, or project one of the systems onto a predetermined state. Nevertheless, the intuition of the controlled case, that parallel free evolution and clever use of \swaps can cause time translations, also holds in our uncontrolled scenario. This is because, for any dimension $d$, it is possible to construct a tensor matrix polynomial in $2$ variables $\Omega(V, W)$ that, evaluated with $d\times d$ matrices $V,W$, is proportional to the \swap operator. We note that, as for the rewinding protocols in the previous section, for some zero-measure sets of $d\times d$ matrix pairs $(V,W)$ the polynomial will vanish. In almost all cases however, the norm of $\Omega(V, W)$ will be finite showing that there is some probability of swapping the states of two uncontrollable systems via scattering experiments. We prove this rather surprising result in \appref{app:PermutationPoly} with a method to construct such a polynomial.

The construction provided in the proof, however, produces polynomials of extraordinarily large degrees which would have very small probabilities of success. For this reason, we also provide, in \appref{app:SWAPNumerical}, a numerical method that, for a given $d$ and $m$, generates all the valid $\Omega(V, W)$. Being a numerical construct, there is no particular structure visible to us in these polynomials or any intuition to be grasped about this result of matrix polynomial theory. To give a sense of scale, we note that the smallest `swap polynomial' $\tilde{\Omega}(V, W)$ that exists in dimension $d=2$ is of degree $5$ and has $40$ terms: the reader can find it in \appref{app:SWAPNumerical}.

Armed with an $\Omega(V, W) \propto \swap$ we are now able to create a fast forwarding protocol for uncontrolled systems. Consider the tensor matrix polynomial
\be
E(V,W)=(V^s\otimes \id)\Omega(V,W)(\id\otimes V^s)\Omega(V,W),
\label{fastforward}
\ee
which is of degree $s+2 \text{deg}(\Omega)$. To show that it evolves the two target systems in the desired way, we directly compute its action:
\begin{align}
\ket{\psi'_{1,2}}&\propto E(V,W)\ket{\psi_{1,2}}\nonumber\\
&\propto (V^s\otimes \id)\mbox{SWAP}(\id\otimes V^s)\mbox{SWAP}\ket{\psi_{1,2}}\nonumber\\
&\propto (V^{2s}\otimes \id)\ket{\psi_{1,2}}. \nonumber
\end{align}
That is, using a scattering protocol with running time $T'=\left(s+2 \mbox{deg}(\Omega)\right)\Delta t$, we can project systems $1,2$ to the state $U(T_1,0)$, with $T_1=2s\Delta t$. As before, by taking $\Delta t$ small enough while keeping $s\Delta t$ constant, we have that $T_1=2T'$, saturating the bound provided by \eqref{general}.

While the probability of success is not the focus of this work, it is still interesting to calculate it for some examples to show that it is not too small. In the qubit case, and taking $V$ and $W$ independently from the Haar measure, we find that the probability of implementing the \swap polynomial $\tilde{\Omega}(V, W)$ mentioned above using the canonical scattering experiment is around $0.7\%$. To find a probability for the corresponding $E(V, W)$ we use the same trick as in the $n=1$ fast forwarding case in order to construct an experiment that has $10$ probes, rather than the canonical $10+s$. In this case we get an average probability of $0.02\%$. While small, these probabilities are not microscopic and, we stress, there is no reason to believe that there do not exist scattering experiments for the same polynomial with much greater success rates.

The remaining task for $n=2$, fast rewinding, is done similarly. Consider the tensor polynomial $D(V,W)\equiv \left(R\otimes \id\right)\Omega\left(\id\otimes R\right)\Omega$, where $R(V,W)\propto V^{-s}$ is the rewinding polynomial used in the $n=1$ case. As a tensor polynomial, $D(V,W)$ has degree $2\mbox{deg}(\Omega)+s(d-1)+d^2$; thus it can be implemented through a scattering protocol with running time $T'=(2\mbox{deg}(\Omega)+s(d-1)+d^2)\Delta t$. Its effect on the joint system $1-2$ is to project it to
\be
D(V,W)\ket{\psi_{1,2}}\propto (V^{-2s}\otimes \id)\ket{\psi_{1,2}}.
\ee
As before, by taking $\Delta t\to 0$ while keeping $s\Delta t$ constant, we achieve our known bound for time translation $\{T_1=-2T'/(d-1),T_2=0\}$.\\

\noindent\emph{Protocols in general $n$}

In the fully general case of $n$ different systems, we need to be able to realize extremal protocols of the form $\{T_j=nT',T_{k\neq j}=0\}$ and $\{T_j=-\frac{nT'}{d-1},T_{k\neq j}=0\}$ in order to reach all solutions to \eqref{general} by using the same sequential argument as before. To do this, we need the same ingredients as in the $n=2$ case with the extra caveat that we require tensor matrix polynomials $\Omega^{(i, j)}(V, W)$ which, evaluated on $d\times d$ matrices, are proportional to the permutation operator that swaps systems $i$ and $j$. These polynomials exist by virtue of the construction presented in \appref{app:PermutationPoly}. With these objects, we are now able to write the matrix polynomials which realize the extremal protocols allowed by \eqref{general}, listed above.

Take the first case where we wish to transfer evolution time from all systems to system $j$ in the forward direction. If we denote $V$ acting on system $k$ and identity on all other systems by $V_k$, then one can verify that the tensor polynomial
\be
E^j(V,W)=V_j^s\prod_{k\not = j}\Omega^{(j,k)}(V, W) V_k^s \Omega^{(j,k)}(V, W),
\ee
(which is analogous to \eqref{fastforward}) is of degree $s+O(1)$. On the other hand, $E^j (V,W)\propto V_j^{ns}$. The corresponding scattering protocol thus evolves system $j$ by $T_j=ns\Delta t$ in $T'= \left(s + O(1)\right)\Delta t$ time units, while keeping the rest of the systems frozen. As before, by taking the limit $\Delta t\to 0$ while keeping $s\Delta t$ constant, we can transfer $nT'$ time units to system $j$ by means of a protocol of duration $T'$. To transfer all the evolution time to one system backwards we need the polynomial 
\be
D^j(V,W)=R_j^s\prod_{k\not = j}\Omega^{(j,k)}(V, W) R_k^s \Omega^{(j,k)}(V, W),
\ee
where $R_j$ is the single particle rewinding protocol from the previous sections. $D^j(V,W)$ has degree $s(d-1) + O(1)$ and is proportional to $V_j^{-ns}$, which saturates the remaining bound of \eqref{general} asymptotically.\\

We note in passing that, as well as the $\Omega^{(i, j)}$ used above, our construction in \appref{app:PermutationPoly} can also produce tensor matrix polynomials which, under evaluation with $d\times d$ matrices, are proportional to a fixed linear combination of permutation operators $\{{\cal P}_\pi:\pi \in S_n\}$, with ${\cal P}_\pi\ket{i_1}...\ket{i_n}=\ket{i_{\pi(1)}}...\ket{i_{\pi(n)}}$. As an additional point we also prove, in \appref{app:InvariantPoly}, that if a tensor polynomial is proportional to the same operator for all matrix variables, then it is proportional to a linear combination of the ${\cal P}_\pi$.

\section*{Conclusion}

In this paper we have characterized how one can probabilistically conduct independent time translations of uncontrolled systems of known dimension by means of scattering experiments. We have seen that, in such scenarios, evolution time behaves like a material resource, in the sense that it can be transferred and destroyed, but not created. It can also be inverted, at a cost, via an irreversible process.  From a wider perspective, making objects age slower or faster than they should, or even rejuvenate them, is a recurring theme in both mythology and science fiction \cite{Gleick2016}. This motivates, if only for historical reasons, our theoretical work, as well as future experimental implementations of the scattering protocols hereby presented.

Our results could also be interpreted as establishing a way of trading off between time and space resources in a computational sense. If the free evolution of a system after a certain time encodes the solution to a computational task, we are able to reduce the time needed by having multiple copies of the system and quickly translating one to the future. The probability of success can even be raised arbitrarily high by taking multiple copies of the ensemble and running the protocol in parallel. While this is unlikely to be a feasible scheme to speed up a practical computer, it may be a useful way to think of converting between time and space complexity.

A key technical point we proved is that one can devise tensor matrix polynomials which, evaluated on $d\times d$ matrices, are proportional to a given permutation operator. As a consequence, we can devise scattering protocols which permute the quantum states of any uncontrolled systems present in the scattering region. Open questions about these polynomials include finding the smallest possible degree required for a given $n$, $d$ and target permutation, and an intuitive way of constructing them. We expect that tools from representation theory are likely to be very productive in tackling these questions. A better understanding of matrix polynomials is also crucial to determine the optimal probability of success of scattering experiments, a question we have only touched on here.

\section*{Acknowledgements}

This work was supported by the Austrian Science Fund (FWF): P 30947. D.T. became a recipient of a DOC Fellowship of the Austrian Academy of Sciences at the Institute for Quantum Optics and Quantum Information (IQOQI) in the closing stages of the work. The ERC did not contribute to the funding of this project.

\bibliographystyle{apsrev4-1}
\bibliography{Manuscript}

\newpage
\appendix
\section*{APPENDICES}

\section{Rewinding Polynomial}
\label{app:RewindingPoly}

In the main body of the text, we showed how time-rewinding with a single system could be done at the optimal rate of $1/(d-1)$ with the use of the matrix polynomial $R(V, \vec{W}) \propto V^{-s}$. We now give a construction for such a polynomial in arbitrary $d$.

We begin by considering a \emph{central matrix polynomial for dimension $d$}; this is a matrix polynomial $G(X_1,...,X_L)$ with no constant term with the property that, evaluated on $d\times d$ matrices, it is always proportional to the identity matrix,
$$
G(X_1,...,X_L) = g(X_1,...,X_L)\id_d,
$$
where the scalar $g$ is non-zero for generic $d\times d$ matrices. It is shown in \cite{formanek} that, for any $d$, there exists a homogeneous central matrix polynomial in $d+1$ variables $F_d(V,Y_1,...,Y_d)$ for dimension $d$ with degree $d^2$ and linear in the variables $Y_1,...,Y_d$.

We now evaluate this polynomial with $Y_i = W_i V^s$ and define the new polynomial $R_d$ according to 
$$
F_d(V, W_1 V^s,..., W_d V^s) = R_d(V, W_1, ..., W_d) V^s \propto \id_d,
$$
where the linearity in the $Y_i$ variables of $F_d$ is used to factor out the $V^s$. From the invertibility of $V$ we have that $R_d(V, W_1, ..., W_d) \propto V^{-s}$.  As $F_d$ has degree $d^2$ in its variables, $R_d$ has degree $s(d-1) + d^2$. This is what we required in the main text to show that there exists scattering experiments that can, asymptotically, do all the time translations allowed by \eqref{general} in the single system case.

\section{Permutation Polynomials}
\label{app:PermutationPoly}

In order to perform any time-translation operations on $n$ systems of dimension $d$ compatible with \eqref{general} with a scattering experiment we required, in the main body of the text, a tensor matrix polynomial proportional to the $\swap$ operator between any two systems. From the targeting assumption, we can clearly restrict ourselves to the $n=2$ case without loss of generality. We now present in detail a constructive method for obtaining such a polynomial, and then show how it extends to matrix polynomial proportional to an arbitrary permutation operator ${\mathcal{P}}_\pi$ in general $n$.

This construction requires a lot of matrix variables; let us start with $X_1,...,X_m$. Note that any linear combination $C$ of the matrices $X_i\otimes X_i$ is such that ${\cal P}_{(1,2)} \cdot C\cdot {\cal P}_{(1,2)}=C$. This implies that $C=C_S\oplus C_A$, where $S$ and $A$ denote, respectively, the symmetric and antisymmetric subspaces of $\C^d\otimes \C^d$, with dimensions $d(d+1)/2$ and $d(d-1)/2$. 

Notice that any central polynomial of dimension $D$ vanishes when evaluated with matrices of dimension $D'<D$. This is so because we can embed any $D'\times D'$ matrix $M$ into $D\times D$ matrices as $M\oplus 0_{D-D'}$. If we evaluate the considered central polynomial with such $D\times D$ matrices, the result should be proportional to the identity. At the same time, it must be of the form $M'\oplus 0_{D-D'}$. It follows that the proportionality constant is zero. 

Now, let $G(Z_1,...,Z_m)$ be a central polynomial for dimension $d(d+1)/2$, and let $C_1(X),...,C_m(X)$ be linear combinations of $X_1\otimes X_1,...,X_m\otimes X_m$ such that, for some $d\times d$ matrices $X_1,...,X_m$, $G(C_1(X),...,C_m(X))\not=0$. Since the dimensionality of the symmetric (antisymmetric) space is equal to (smaller than) $d(d+1)/2$, by the previous observations we have that $\tilde{G}(X)\equiv G(C_1(X),...,C_m(X))$ must be proportional to $\id_S\oplus 0_A$ when evaluated with $d\times d$ matrices. That is, the polynomial $\tilde{G}(X)$ is generically nonzero and proportional to the symmetric projector $\Pi_S$.

Given another set of matrix variables $\{Y_i\}_i$, consider the polynomials $P_{ij}(Y)\equiv Y_i\otimes Y_j-Y_j\otimes Y_i$. Any linear combination $C$ of those satisfies ${\cal P}_{(1,2)}\cdot C\cdot {\cal P}_{(1,2)}=-C$. This implies that 

\be
C=\left(\begin{array}{cc}0_S&C_1\\C_2&0_A\end{array}\right).
\ee

Therefore, the image of the polynomials $\tilde{P}_{ijkl}\equiv P_{ij}\tilde{G}P_{kl}$, when evaluated on $d\times d$ matrices, are matrices of the form $0_S\oplus \bullet_A$. In turn, any matrix of the form $0_S\oplus \bullet_A$ can be expressed as a linear combination of $P_{ijkl}(Y)$, for certain $Y's$. Now, choose a central matrix polynomial $H(Z_1,...,Z_{m'})$ for dimension $d(d-1)/2$, and choose linear combinations $\tilde{C}_1,...,\tilde{C}_{m'}$ of the polynomials $\tilde{P}_{ijkl}(X,Y)$ such that $\tilde{H}(X,Y)\equiv H(\tilde{C}_1(X,Y),...,\tilde{C}_{m'}(X,Y))\not=0$, for some $X,Y$. It follows that $\tilde{H}(X,Y)$ is generically nonzero, and proportional to the antisymmetric projector $\Pi_A$.

The problem is that the proportionality scalars in $\tilde{G}$ and $\tilde{H}$ may differ. To fix this, choose a central matrix polynomial $F(Z_1,...,Z_p)$ for dimension $d^2$, and let $Q_1=T_1\otimes T_2, Q_2=T_3\otimes T_4...,Q_p=T_{2p-1}\otimes T_{2p}$, where $T_1,...,T_{2p}$ are new matrix variables. Then we have that

\be
\tilde{F}=F\left((\tilde{G}+\tilde{H})Q_1(\tilde{G}+\tilde{H}),...,(\tilde{G}+\tilde{H})Q_p(\tilde{G}+\tilde{H})\right)
\ee
\noindent is a central polynomial. We can express it as 

\be
\tilde{F}=\tilde{G}F^1\tilde{G}+\tilde{G}F^2\tilde{H}+\tilde{H}F^3\tilde{G}+\tilde{H}F^4\tilde{H},
\ee
\noindent for some polynomials $F^1,F^2,F^3,F^4$. 

Let $f(X,Y,T)$ be the scalar function satisfying $\tilde{F}(X,Y,T)=f(X,Y,T)\id_{d^2}$. Then we have that

\begin{align}
&f(X,Y,T)\Pi_S=\Pi_S\tilde{F}(X,Y,T)\Pi_S=\tilde{G}F^1\tilde{G},\nonumber\\
&f(X,Y,T)\Pi_A=\Pi_A\tilde{F}(X,Y,T)\Pi_A=\tilde{H}F^4\tilde{H}.
\end{align}

\noindent The two polynomials we are looking for are thus $\tilde{S}\equiv \tilde{G}F^1\tilde{G},\tilde{A}\equiv \tilde{H}F^4\tilde{H}$. Since $\id=\Pi_S+\Pi_A$, ${\cal P}_{(1,2)}=\Pi_S-\Pi_A$, by combining the polynomials $\tilde{S},\tilde{G}$, we can induce any linear combination of the two permutation operators on $n=2$ systems. 

There is, though, a subtlety. The way they were constructed, $\tilde{S},\tilde{A}$ are not homogeneous: the reason is that $\tilde{G},\tilde{H}$ have different degree. This can be fixed, e.g., by redefining them as $\tilde{G}\to \tilde{G}J_G,\tilde{H}\to \tilde{H}J_H$, where $J_G,J_H$ are central polynomials for dimension $d^2$ on new variables such that $\mbox{deg}(\tilde{G})+\mbox{deg}(J_G)=\mbox{deg}(\tilde{H})+\mbox{deg}(J_H)$. Such polynomials always exist: indeed, for all dimensions there exists a central matrix polynomial of the form $J(U_1,...,U_q)$, linear in $U_1$ \cite{formanek}. Its degree can be increased by an arbitrary amount $k$ via the transformation $J(U_1,...,U_q)\to J(U^{k+1}_1,...,U_q)$.

It remains to show that one can extend this construction for general $n$. This follows from the fact that any permutation of $n$ parties can be expressed as the product of $n-1$ gates which are either permutations of the form $(j,k)$ or identities. Define the tensor polynomial $P^{ij}$ as $P^{ij}=(\tilde{S}+(-1)^{\delta_{ij}}\tilde{A})_{ij}\otimes \bigotimes_{k\not=j,i} C_k$, where $C$ is a central polynomial of the same degree as $\tilde{S},\tilde{A}$. Then, 

\begin{align}
P^{ij}=&\lambda \id, \mbox{ for } i=j;\nonumber\\
&\lambda {\cal P}_{(i,j)},\mbox{ otherwise}.
\end{align}

Now, given a permutation $\pi\in S_n$, define the sequence of permutations $\pi_k=(k,\pi_{k-1}(k))\circ\pi_{k-1}$, with $\pi_0=\pi$. It follows that $\prod_{k=1}^{n-1}P^{k,\pi_{k-1}(k)}=\lambda^{n-1}{\cal P}_{\pi}$.

\section{Invariant Tensor Polynomials}
\label{app:InvariantPoly}

In the previous section we showed how to construct a tensor matrix polynomial $G(X_1, ..., X_L)$ such that it was proportional to a permutation operator between the identical systems. As the proportionality constant is the same for all permutations, this allows arbitrary linear combinations of permutations to be realised independently of $\{X_i\}_i$. We now show that these are the only invariant tensor matrix polynomials.

 Let us consider a general such polynomial $G(X_1, ..., X_L) \propto M$. We require that it remain proportional to the same $M$ for different variables, which includes the unitary rotation $G(U X_1 U^\dagger, ..., U X_L U^\dagger)$ for $U\in SU(d)$. From the tensor product structure of $G$ this implies
 \be
 U^{\otimes n} M (U^{\dagger})^{\otimes n} = \lambda_U M,
 \ee
where $\lambda_U$ is a constant of proportionality and the equation must hold for all $U$. Vectorising this equation by applying the transformation 
\be
M = \sum_{i, j} m_{ij} \ket{i}\bra{j} \to \sum_{i, j} m_{ij} \ket{i}\ket{j} = \ket{M},
\ee
leads to
\be
\label{eqApp:Eigenvalue}
 U^{\otimes n}\otimes(U^*)^{\otimes n} \ket{M}= \lambda_U \ket{M}.
\ee
From the Peter-Weyl theorem of representation theory \cite{Knapp1986}, we decompose the group action into a direct sum of irreducible representations,
\begin{align}
U^{\otimes n}\otimes(U^*)^{\otimes n} &= \bigoplus_i r_{d_i}^{\oplus k_i} \\
\ket{M} &= \bigoplus_i \bigoplus_{j=1}^{k_i} \ket{m_{d_i}^j},
\end{align}
where $r_{d_i} \in SU(d_i)$ and $k_i$ is the multiplicity. The equality in \eqref{eqApp:Eigenvalue} must hold over all tensor blocks independently which implies
\be
r_{d_i} \ket{m_{d_i}^j} = \lambda_U \ket{m_{d_i}^j} \quad \forall \.i, j, U.
\ee
For $d_i > 1$ this requires that $\lambda_U = 0$ as there is no non-trivial eigenvector for every matrix in $SU(d_i)$. However, for $d_i = 1$ then $r_{d_1} = 1$, and therefore  $\lambda_U = 1$ for all $\ket{m_1^j}$. Hence, the only $\ket{M}$ which can satisfy \eqref{eqApp:Eigenvalue} are those which lie in the span of $\ket{m_1^j}$, and they do so with $\lambda_U=1$. This means that $M$ lies in the span of the trivial representation of the conjugate action of $U^{\otimes n}$. By the Schur-Weyl duality, this is exactly the same space as the permutations $S_n$.

\section{Finding \swap polynomials numerically}
\label{app:SWAPNumerical}

The numerical method used to find matrix polynomials for the \swap operation between two systems is similar to the one described in \cite{resetting} for central polynomials. First, we write a general tensor matrix polynomial as
\be
P(X_1, ..., X_{D}) = \sum_{\vec{\imath}, \vec{\jmath}} p_{\vec{\imath}, \vec{\jmath}}\;X_{i_1} X_{i_2} ... X_{i_m} \otimes X_{j_1} X_{j_2} ... X_{j_m},
\ee
{\miguel where $m$ is the degree of the polynomial; the $X's$ are the matrix variables; $D$ is the number of variables of the desired matrix tensor polynomial, and $\vec{\imath}, \vec{\jmath} = \{1, 2, ..., D\}^{\times m}$. The coefficients $\{p_{\vec{\imath}, \vec{\jmath}}:\vec{\imath}, \vec{\jmath}\}$ can be regarded as entries in the vector space of polynomials $\mathbb{P}$, where a polynomial is expressed as $
 \sum_{\vec{\imath}, \vec{\jmath}} p_{\vec{\imath}, \vec{\jmath}} \ket{i_1,j_1}...\ket{i_m,j_m}$.
 
Our aim is to characterize the subspace of $\mathbb{P}$ corresponding to the set of \swap polynomials. To do so, we first consider the problem of characterizing the subspace $V^\perp$ spanned by all polynomials $P_{X,L,R}$ with coefficients given by

\be
p_{\vec{\imath}, \vec{\jmath}} = \braket{L | \swap \cdot (X_{i_1} X_{i_2} ... X_{i_m} \otimes X_{j_1} X_{j_2} ... X_{j_m}) | R},
\label{random}
\ee
\noindent where $X=(X_1,...X_D)$ is a tuple of $d\times d$ matrices and $\ket{R},\ket{L}\in \C^{d}\otimes\C^d$ satisfy $\ket{R}\perp\ket{L}$.

This can be done by generating sequences of random matrices and vectors $(X^{(j)}, R^{(j)}, L^{(j)})$ and then computing the corresponding polynomials $(P^{(j)})_j$ via the equation above. While we generate $(P^{(j)})_j$, we use Gram-Schmidt to produce a sequence of orthonormal vectors $(\tilde{P}^{(j)})_j$. The procedure ends when the last random polynomial $P^{(N+1)}$ generated is a linear combination of $(\tilde{P}^{(j)})_{j=1}^N$. It can then be proven that, with probability $1$, $V^{\perp}$ is spanned by the vectors $(\tilde{P}^{(j)})_{j=1}^N$.

Now, consider the orthogonal complement of $V^{\perp}$, call it $V$. By construction, for any polynomial $p$ with coefficients $\sum p_{\vec{\imath}, \vec{\jmath}} \ket{i_1,j_1}...\ket{i_m,j_m}\in V$, and any polynomial $P_{X,R,L}$ (with $\ket{R}\perp\ket{L}$), 
\begin{align}
&0=\braket{p, P_{X,R,L}}=\nonumber\\
&\bra{L} \swap \cdot \sum_{\vec{\imath}, \vec{\jmath}} p_{\vec{\imath}, \vec{\jmath}} X_{i_1} ... X_{i_m} \otimes X_{j_1} ... X_{j_m}\ket{R} = \nonumber\\
&=\bra{L}\swap \cdot p(X)\ket{R}.
\label{ortho}
\end{align}
\noindent Fixing $X$ and varying over the vectors $\ket{R}\perp\ket{L}$, we have that $\swap \cdot p(X)\ket{\psi}= \lambda_{\psi}\ket{\psi}$ for all $\ket{\psi}$. It is easy to see that, for this relation to hold for all vectors $\psi$, the proportionality constant $\lambda_{\psi}$ must be independent of $\psi$. It thus follows that $\swap \cdot p(X)\propto\id$ and hence $p(X)\propto \swap$.

Note that there might exist elements of $V$ whose associated tensor polynomials, evaluated on $d\times d$ matrices, always return the zero matrix (which is also proportional to the \swap operator). To subtract those, we similarly characterize the space $N\subset V$ of such ``null polynomials'' by allowing $\ket{L},\ket{R}$ to be non-orthogonal in the above scheme. A basis of linearly independent non-zero swap polynomials is obtained by considering the quotient vector space $V/N$, i.e., the set of vectors in $V$ which are orthogonal to $N$.

For $D = d = 2$, $V/N$ is just the null vector for $m < 5$ and has dimension $3$ for $m=5$. Hence three independent \swap polynomials of degree $5$ can be easily extracted. By taking linear combinations of an element of $V/N$ with vectors in $N$ we can construct non-null swap polynomials with a minimum number of non-zero terms. Using this scheme, the simplest swap polynomial we could construct has $2$ different matrix variables, degree $5$, and $40$ non-zero coefficients:}
\begin{align*}
&\tilde{\Omega}(V,W) = \\
 & VWWVW\otimes VWWVW - VWWVW \otimes WWVVW \\
- & VWWWV \otimes VWWVW + VWWWV \otimes VWWWV \\
+ & VWWWV\otimes WVWVW - VWWWV\otimes WWVWV\\
 - &VWWWW\otimes VWVWV + VWWWW\otimes WVVWV \\
 - & WVWWV\otimes VWWVW - WVWWV\otimes VWWWV \\
 + & WVWWV\otimes WVWVW + WVWWV\otimes WWWVV\\
 -& WVWWW\otimes VWVWV + WVWWW\otimes VWWVV\\
 + & WVWWW\otimes WVWVV - WVWWW\otimes WWVVV\\
 + & WWVWV\otimes VWWVW - WWVWV\otimes VWWWV \\
 - & WWVWV\otimes WVWVW + WWVWV\otimes WVWWV \\
 - & WWVWW\otimes VWVVW + WWVWW\otimes VWVWV\\
 + & WWVWW\otimes WVVVW - WWVWW\otimes WVVWV\\
 +&  WWWVV\otimes VWWWV - WWWVV\otimes WVWVW\\
 + & WWWVV\otimes WWVVW - WWWVV\otimes WWWVV \\
 +&  WWWVW\otimes VWVWV - WWWVW\otimes VWWVV \\
 - & WWWVW\otimes WVWVV + WWWVW\otimes WWVVV \\
 - & WWWWV\otimes VVWWV + WWWWV\otimes VWVVW \\
 +& WWWWV\otimes VWVWV - WWWWV\otimes WVVVW \\
 - & WWWWV\otimes WVWVV + WWWWV\otimes WWVVV \\
 + & WWWWW\otimes VVWVV - WWWWW\otimes VWVVV.
\end{align*}

We conclude this section with an observation on the computational complexity of identifying the vector space $V/N$. For clarity, we achieved this above by characterizing the subspaces $V,N\subset \mathbb{P}$ independently and then computing their quotient. While this procedure works well for $d=2, 3$ and $m\approx15$ or less on a laptop, it is extremely costly for high degree $m$, as it requires one to deal with vectors of dimension exponential in $m$. 

However, there is an alternative method to compute $V/N$, namely, characterizing $N^\perp$ and then finding the largest subspace of $N^\perp$ all whose vectors are contained in $V$. As we next see, this can be achieved in time polynomial in $m$. 

First, note that the vectors in \eqref{random} generating $V^\perp$ and $N^\perp$ can be interpreted as $m$-partite matrix product states (MPS) \cite{tensor_networks} of physical dimension $D^2$ and bond dimension $d^2$. Indeed, the vector $\sum_{\vec{\imath}, \vec{\jmath}}p_{\vec{\imath}, \vec{\jmath}}\ket{i_1,j_1}...\ket{i_m,j_m}$, with components $p_{\vec{\imath}, \vec{\jmath}}$ defined by \eqref{random}, can be expressed as

\be
\sum_{\vec{\imath}, \vec{\jmath}}\braket{\tilde{L} | Y_{i_1,j_1}...Y_{i_m,j_m} | R}\ket{i_1,j_1}...\ket{i_n,j_n},
\label{vector_MPS}
\ee
\noindent with $Y_{i,j} =X_{i}\otimes X_{j}$ and $\ket{\tilde{L}}=\swap\ket{L}$.
This means that we can derive a basis for $N^\perp$ just by generating random MPS and applying the Gram-Schmidt process. Since the scalar product of two $m$-partite MPS can be computed with $O(m)$ operations \cite{tensor_networks}, this basis can be computed efficiently, as long as $\mbox{dim}\left(N^\perp\right)=poly(m)$. To see that this is the case, we extend the argument used in \cite{werner} and \cite{bond_dimension} to lower bound the dimensionality of matrix polynomial identities.

Let the matrices $X_1,...,X_D$ in \eqref{vector_MPS} be $d\times d$, and let $M$ ($\hat{M}$) be the set of all monomials of degree $2m$ ($2$) of the entries of $X_1,...,X_D$ (the entries of $L$ and $R$, and linear in both $L$ and $R$). Then we have that \eqref{vector_MPS} can be expressed as

\be
\sum_{\mu\in M,\hat{\mu}\in \hat{M}}\mu(X)\hat{\mu}(L,R)\ket{\psi_{\mu,\hat{\mu}}},
\ee
\noindent where, for each $\mu\in M,\hat{\mu}\in \hat{M}$, $\ket{\psi_{\mu,\hat{\mu}}}$ is an $m$-partite vector independent of the values of the entries of $X, L, R$.

By varying $X, L, R$, we obtain different linear combinations of the vectors $\{\ket{\psi_{\mu,\hat{\mu}}}:\mu\in M, \hat{\mu}\in\hat{M}\}$. Such linear combinations span $N^\perp$. $N^\perp$ thus has dimension at most $|M||\hat{M}|$, i.e.,

\be
\mbox{dim}(N^\perp) \leq \left(\begin{array}{c}2m + Dd^2 -1\\Dd^2-1\end{array}\right)d^2=O\left(m^{Dd^2-1}\right).
\ee

Similarly, one can find a basis for $V^\perp\subset N^\perp$ by generating random MPS of the form \eqref{vector_MPS}, with $\ket{L}\perp \ket{R}$. Hence, in time polynomial in $m$, we can construct MPS bases ${\cal B}_V$, ${\cal B}_N$ for $V^\perp$, $N^\perp$ with polynomially many elements. Given ${\cal B}_V$, ${\cal B}_N$, through basic linear algebra one can produce a basis for the subspace $V/N=\{v:v\in N^{\perp}, v\perp V^{\perp}\}$, with elements expressible as linear combinations of polynomially many MPS (more precisely, the elements of ${\cal B}_N$). Moreover, inasmuch as we can efficiently compute the overlaps between any two vectors $p, v\in {\cal B}_V\cup {\cal B}_N$, this can be achieved in $poly(m)$ time. 

\end{document}